\pdfoutput=1
\documentclass[twocolumn]{svjour3}
\usepackage[utf8]{inputenc}                               
\usepackage[english]{babel}  
\usepackage[T1]{fontenc}                                      
\usepackage{hyperref}
\usepackage{graphicx}
\usepackage{siunitx}
\usepackage{amssymb}
\usepackage{gensymb}
\usepackage{mathtools}
\def\makeheadbox{{%
		\hbox to0pt{\vbox{\baselineskip=10dd\hrule\hbox
				to\hsize{\vrule\kern3pt\vbox{\kern3pt
						\hbox{This is a pre-print of an article published in Applied Physics B. }
						\hbox{The final authenticated version is available online at: \href{https://doi.org/10.1007/s00340-020-07424-5}{https://doi.org/10.1007/s00340-020-07424-5}.}
						\kern3pt}\hfil\kern3pt\vrule}\hrule}%
			\hss}}}

\DeclarePairedDelimiter\ket{\lvert}{\rangle}
\DeclarePairedDelimiterX\braket[2]{\langle}{\rangle}{#1 \delimsize\vert #2}

\usepackage[numbers,sort&compress]{natbib}

\begin{document}
	\title{Measurement and Simulation of Atomic Motion in Nanoscale Optical Trapping Potentials}               
	\date{January 31, 2020}             
	\author{Signe B. Markussen \and J{\"u}rgen Appel \and Christoffer {\O}stfeldt \and Jean-Baptiste S. B{\'e}guin \and Eugene S. Polzik \and J{\"o}rg H. M{\"u}ller}

	\institute{Signe B. Markussen \and J{\"u}rgen Appel \and Christoffer {\O}stfeldt \and Jean-Baptiste S. B{\'e}guin \and Eugene S. Polzik \and J{\"o}rg H. M{\"u}ller \at
		 QUANTOP, Niels Bohr Institute, University of Copenhagen, Blegdamsvej 17, 2100 Copenhagen, Denmark \\ Tel.: +45 35 32 53 04, Fax: +45 35 32 50 16 \\ \email{muller@nbi.ku.dk}
	\and
	J{\"u}rgen Appel \at Danish National Metrology Institute, Kogle All{\'e} 5, 2970 H{\o}rsholm, Denmark 
	\and
	Jean-Baptiste S. B{\'e}guin \at Norman Bridge Laboratory of Physics, Caltech, Pasadena, CA 91125, U.S.
	}
		
	\maketitle
	
	\begin{abstract}
		Atoms trapped in the evanescent field around a nanofiber experience strong coupling to the light guided in the fiber mode. However, due to the intrinsically strong positional dependence of the coupling, thermal motion of the ensemble limits the use of nanofiber trapped atoms for some quantum tasks. We investigate the thermal dynamics of such an ensemble by using short light pulses to make a spatially inhomogeneous population transfer between atomic states. As we monitor the wave packet of atoms created by this scheme, we find a damped oscillatory behavior which we attribute to sloshing and dispersion of the atoms. Oscillation frequencies range around $\SI{100}{\kilo\hertz}$, and motional dephasing between atoms happens on a timescale of $\SI{10}{\micro\second}$. Comparison to Monte Carlo simulations of an ensemble of $1000$ classical particles yields reasonable agreement for simulated ensemble temperatures between $\SI{25}{\micro\kelvin}$ and $\SI{40}{\micro\kelvin}$.
	\end{abstract}

	\section{Introduction}
		Efficient coupling of atomic ensembles to well defined propagating optical modes is an important theme in quantum optics. While for a single atom or ion a good overlap between a controlled optical mode and the dipole emission pattern of the particle can be achieved with high numerical aperture lens systems, it is less obvious how to achieve similarly strong coupling for atomic ensembles with a limit on particle density. Over the last decade several experimental groups developed nanofiber traps, where atomic ensembles are trapped in the evanescent part of light modes guided by an optical fiber carefully thinned to subwavelength diameter \cite{Kien2004,Vetsch2010,Lacroute2012,Beguin2014,Gouraud2015,Kato2015,Li2017,Ruddell2017,Solano2017}. Here, a focus condition for light is maintained over the whole length of the nanofiber waist and the strong intensity gradients in the evanescent field outside the fiber can be harnessed to sculpt an extreme 1D dipole potential landscape for trapping atomic ensembles with far detuned laser light.
		The coupling of atoms to near resonant light through the fiber, characterized by the optical depth per atom, can reach values of ${OD=10\%}$ at the trap minima for well chosen fiber diameters and trapping wavelengths \cite{Vetsch2010}. 
		
		Intrinsically, the coupling varies strongly with both the distance of the atoms from the surface of the nanofiber and with their azimuthal position with respect to the polarization of the probe light mode. This entails classical noise in the coupling of the ensemble as individual atoms move thermally in the trapping potential at any non-zero ensemble temperature. The statistics and correlations of this coupling noise put restrictions on the use of nanofiber trapped ensembles for quantum tasks, e.g. quantum memories or heralded single photon sources \cite{Gouraud2015,LeKien2016,Beguin2018,Kato2019}, and provide a strong incentive to develop methods to cool the atoms as close as possible to the motional ground state of the trapping potential \cite{Albrecht2016,Ostfeldt2017,Meng2018}.
		
		To characterize the limitations, it becomes interesting to investigate experimentally the intrinsic thermal motion in atomic ensembles trapped around a nanofiber. 
		To this end, we use short light pulses to make a spatially inhomogeneous Raman population transfer between two stable electronic ground states of atoms. This creates a wave packet of ``tagged'' atoms close to the fiber; we then study the evolution of this wave packet as time progresses. As the atoms move thermally, this packet of atoms will slosh and spread in the trap potential, and monitoring this will yield information about the trap frequencies and motional dephasing of the atoms. This experimental strategy is related to the echo spectroscopy technique presented in \cite{Oblak2008}. However, the faster motional time scales for nano-scale atom traps require us to work at much higher Rabi frequencies than is currently feasible for microwave echo spectroscopy.
		
		Some care must be taken to not perturb the ensemble while attempting the measurements, as this would create extra motion that can not be attributed to the thermal fluctuations. For this reason, the light fields used for the population transfer are balanced in a way that cancels out the differential and common mode Stark shifts effected by the light\cite{Ostfeldt2017}, such that intensity gradients do not cause classical dipole forces.

		To gain additional understanding of the thermal motion, we run simulations of classical particles moving in the trap potential. From the trajectory simulations we can construct observables that correspond to the ones we can extract from the experimental data. We find reasonable quantitative agreement between measured and simulated signals and can thus use the simulations to determine parameters as the ensemble temperature from the data.
	\begin{figure}
		\centering
		\includegraphics{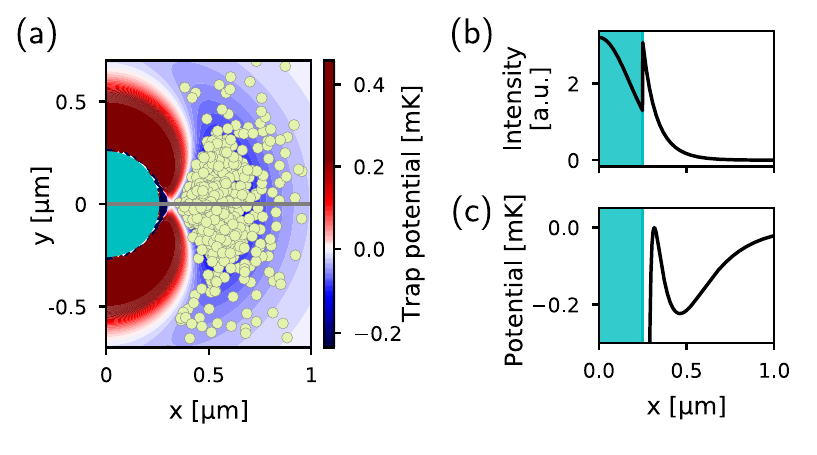}
		\caption{(a) Trap potential around the fiber. Blue (red) indicates negative (positive) potential energy. A $\SI{30}{\micro\kelvin}$ ensemble of non-interacting atoms can be seen as yellow dots. In experiments, a similar ensemble exists on the other side of the fiber. (b) Intensity variation of probe light along the x-direction (grey line in (a)). (c) Radial trapping potential, cut along the x-direction. Teal areas are inside the nanofiber. }
		\label{fig:potential}
	\end{figure}
	\section{Methods and Results}
		We trap atoms in the evanescent field of a standard optical fiber tapered down with linear tapers to a nanosection with a diameter of ${\varnothing = \SI{470}{\nano\meter}}$. We create a dipole trap for cesium (Cs) atoms using two counter-propagating beams of red-detuned light ($\lambda = \SI{1056}{\nano\meter}$, $P\simeq2 \times \SI{1.5}{\milli\watt}$) and one beam of blue-detuned light ($\lambda = \SI{780}{\nano\meter}$, ${P\simeq\SI{8}{\milli\watt}}$) with the orthogonal quasi-linear polarization. The resulting potential forms two 1-dimensional strings of trapping sites, one on each side of the fiber, where atoms can be trapped in the anti-nodes of the red standing wave. A transversal cross section of the resulting trapping potential at such an anti-node can be seen in \autoref{fig:potential} for one side of the fiber. Trap parameters are such that we are within the collisional blockade regime, where each site is occupied by at most one atom, which means that the trapped atoms can be regarded as non-interacting.
		
		The atoms are loaded into the trap using a 6-beam magneto-optical trap (MOT) followed by sub-doppler optical molasses cooling. For the remainder of the experimental sequence, a magnetic bias field of ${B=\SI{3}{G}}$ is applied, leading to a frequency splitting of around $\SI{1}{\mega\hertz}$ between adjacent $m_F$ states. Atoms are then prepared in the state $\ket{{F=3},{m_F=0}}$ by optical pumping before the measurements are performed. A typical ensemble comprises $N\simeq1000$ atoms.
		
		For detection of the atoms, we employ a dispersive probing scheme as described in \cite{Beguin2018}, using two co-propagating probe beams going through the fiber, detuned by $\pm \SI{62.5}{\mega\hertz}$ from the $\ket{{F=4}}\rightarrow \ket{{F'=5}}$ transition of the Cs principal line. We detect the differential phase shift on the two beams using a homodyne measurement with an external local oscillator tuned to the average of the two probe frequencies and an appropriately chosen local oscillator phase. The detected differential phase shift is proportional to the number of atoms in the $\ket{F=4}$ state, with a proportionality factor which only weakly depends on the ensemble temperature. 
		
\begin{figure}
		\centering
		\includegraphics{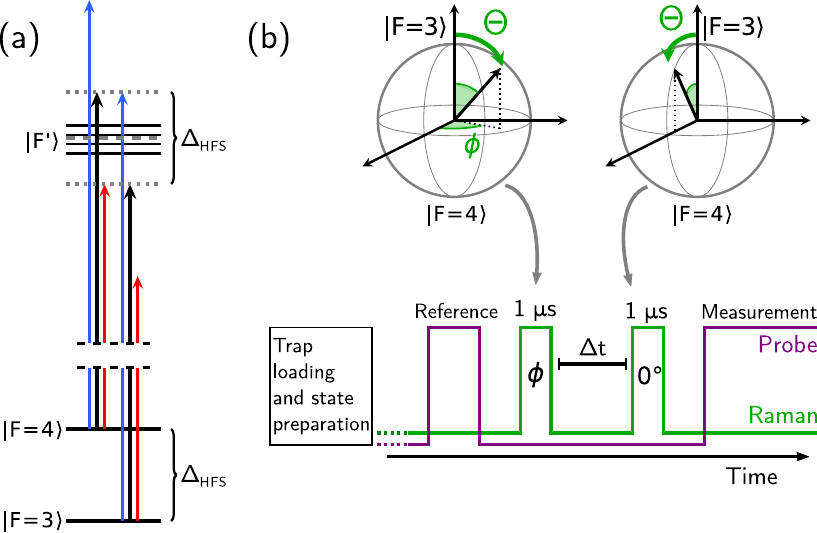}
		\caption{(a) Simplified three-color scheme for dipole-force free Raman transitions. See main text for further explanation of the scheme. (b) Experimental sequence. The atoms are loaded and prepared in $\ket{{F=3},{m_F=0}}$. A $\SI{1}{\micro\second}$ pulse with phase of the modulation tone $\phi$ prepares a wave packet by making a position-dependent pulse angle $\Theta$. After a waiting time $\Delta t$, another $\SI{1}{\micro\second}$ pulse with phase $0$ is applied to the atoms, and finally the probe measures the population of $\ket{F=4}$. This creates sinusoidal fringes as can be seen in \autoref{fig:ramanfringes}. }
		\label{fig:taggingsequence}		
	\end{figure}

		Due to the strong spatial gradient of any light field that propagates through the nanofiber (see \autoref{fig:potential}(b)), it is important that the light used to monitor the atomic motion does not itself exert a force on the atoms. In \cite{Ostfeldt2017}, a dipole-force free scheme for driving Raman transitions in a nanofiber trap is introduced. We use a single phase-modulated beam, co-propagating with the probe light, to drive coherently Raman transitions. The modulation tone is tuned to the ground-state hyperfine splitting of cesium, creating sidebands at ${\Delta= \pm \Delta_{\mathrm{HFS}}}$, corresponding to a two-photon detuning of ${\delta=0}$. See \autoref{fig:taggingsequence}(a) for an illustration of the scheme. 	
		Using microwave Ramsey spectroscopy \cite{Albrecht2016} as a monitoring tool, it is possible to adjust the optical carrier frequency and the carrier-to-sideband ratio such that the dipole forces caused by the various frequency components of the beam cancel out, and no net dipole force is exerted onto the atoms. With this technique, we can monitor the thermal motion of the atoms in a minimally invasive way, without perturbing the original trap potential.
	
	In a single experimental run, a Raman-Ramsey pulse sequence is employed. We use two pulses of light each with a duration of $\SI{1}{\micro\second}$, much shorter than the motional period of the atoms. The choice for this duration is a compromise between gaining a good spatial resolution for the wave packet, while avoiding to drive Raman transitions to adjacent $m_F$ levels by the Fourier broadened spectrum of the pulses.
	
	The first pulse creates some small population transfer from $\ket{{F=3}}$ to $\ket{{F=4}}$ with a pulse angle $\Theta(r)$ that is dependent on the radial and azimuthal position of the atom. This creates a wave packet of atoms close to the fiber that have a higher probability of transfer to the $\ket{{F=4}}$ state than atoms further away from the fiber surface. We then let the wave packet slosh and disperse in the trap potential for a waiting time $\Delta t$, and afterwards attempt to transfer atoms back to $\ket{{F=3}}$ using a similar second light pulse, now with the phase $\phi$ of the modulation tone shifted by $180\degree$. For an idealized case of immobile atoms, the second pulse completely undoes the population transfer made by the first pulse. However, for a finite temperature ensemble, atoms have moved to a new constellation during the waiting time, the strength of the coupling between the atoms and the inhomogeneous light field has changed, and the second pulse is unable to completely undo the first population transfer.
	
	\begin{figure}
		\centering
		\includegraphics{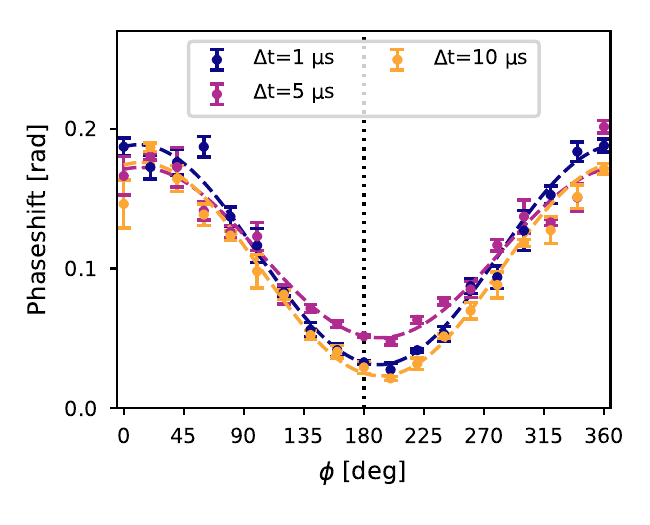}
		\caption{Sinusoidal fringes recorded using the scheme seen in \autoref{fig:taggingsequence} for three waiting times $\Delta t$. The atomic phase shift signal measured in radians is proportional to the number of atoms in $\ket{F=4}$. Points are averages over three measurements, errorbars derived from the standard deviation. Dashed lines are fits according to the model in \autoref{eq:fringemodel}. Note that the bottom level of the fringe (around $\phi=180\degree$) changes with waiting time. This indicates the movement of the wave packet of atoms.}
		\label{fig:ramanfringes}		
	\end{figure}
	
	There is no inherent reason to limit the pulses to be $180\degree$ apart in phase; repeating the sequence and scanning the phase between the two pulses gives additional information. The timing scheme for an experimental run is shown in \autoref{fig:taggingsequence}(b). The first pulse of light has a phase $\phi$, while the second pulse has a fixed phase of $0\degree$. Measuring the net population transfer as a function of the phase difference $\phi$ produces a sinusoidal fringe. Examples of fringes for various waiting times are displayed in \autoref{fig:ramanfringes}. Note the behavior of the bottom part of the fringe, for phases around $\phi\sim 180\degree$. We see that the bottom level of the fringe starts to rise, and then fall again with increasing $\Delta t$, indicating motion of the atoms. We remark that fringe contrast and offset in an equivalent Ramsey experiment with spatially homogeneous microwave pulse excitation evolves smoothly and more than an order of magnitude slower.
	
\begin{figure}
		\centering
		\includegraphics{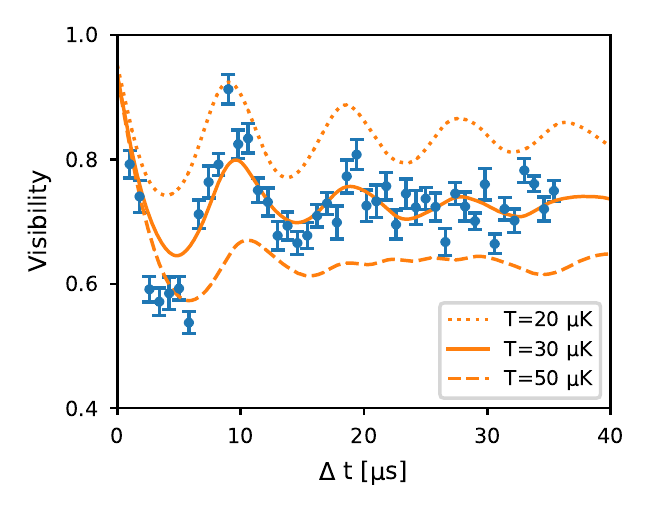}
		\caption{Fringe visibility $V$ as a function of waiting time for the data (dots) and the simulated particles for three ensemble temperatures (dotted, full and dashed lines). Data and corresponding simulations exhibit similar damped oscillations due to the motion and dephasing of the atoms.}
		\label{fig:result_exp_and_sim}
	\end{figure}

	In order to quantify the fringe behavior, we fit the fringes to a model function
	\begin{align}
	f(\phi) = A\cdot \cos(\phi-\phi_0)+k,\label{eq:fringemodel}
	\end{align}
	where $A$ is the amplitude, $\phi_0$ is a phase and $k$ is an offset. We then define the fringe visibility $V=A/k$, and plot it as a function of waiting time. This can be seen in \autoref{fig:result_exp_and_sim}, where a clear damped oscillatory behavior with waiting time is evident. The total power of the Raman laser light employed in this experimental run is $P=\SI{26.3(4)}{\nano\watt}$. The choice of power and duration for the transfer pulses allows to adjust the pulse angle distribution and hence the size and shape of the prepared wave packet. Experimental runs with pulse powers between $\SI{24}{\nano\watt}$ and $\SI{42}{\nano\watt}$ display similar behavior of the visibility with waiting time. 
	
	We calculate that for the highest used pulse power of $\SI{42}{\nano\watt}$, the pulse angle for an atom at the trap bottom amounts to $\Theta(\SI{0.46}{\micro\meter}) \simeq \SI{1.25}{\radian}$. In comparison, atoms located at the top of the repulsive barrier experience a pulse angle of $\Theta(\SI{0.38}{\micro\meter}) \simeq \SI{2.9}{\radian}$. While in principle higher pulse powers can be applied, this would lead to non-trivial shapes of the wave packet. 
	
	Fitting the extracted visibility data to a damped harmonic oscillator model (fit not shown in \autoref{fig:result_exp_and_sim}), we find that the frequencies of oscillation for different data sets range between $\SI{90}{\kilo\hertz}$ and $\SI{110}{\kilo\hertz}$. This is in good agreement with a trap frequency of around $\SI{100}{\kilo \hertz}$, expected from calculating the trapping potential from the fiber diameter and trap powers.
	The time constant for the damping ranges between $\SI{8}{\micro\second}$ and $\SI{14}{\micro\second}$. We attribute this damping to motional dephasing between the atoms due to the anharmonicity of the trap potential. In fact, different from the textbook case of a harmonic oscillator potential, revivals of the spatial atomic configuration are never complete in an anharmonic and nonseparable potential landscape.
	
	\section{Comparison to trajectory simulations}
	
\begin{figure}
		\centering
		\includegraphics{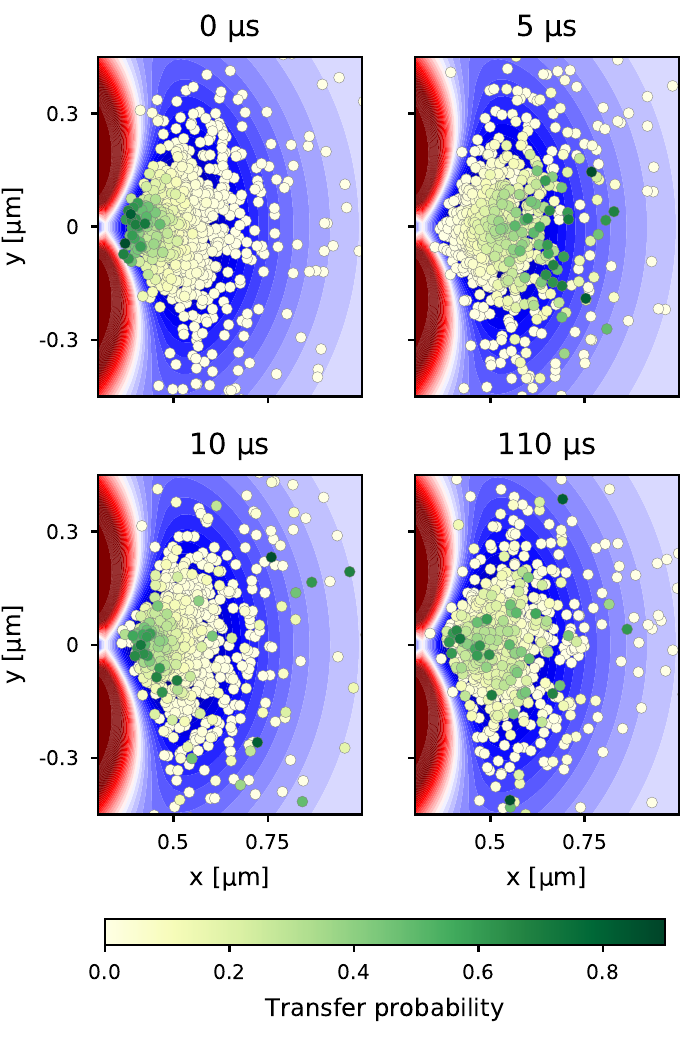}
		\caption{Snapshots of the simulated ensemble of classical particles moving in the trap potential at different times after the end of a $\SI{1}{\micro\second}$ pulse of light, making a population transfer on a packet of atoms close to the fiber. Each particle (dots) is colored according to the probability of the particle being in $\ket{F=4}$. As time progresses, the shape of the packet gets less defined, and eventually washes out. This particular ensemble is simulated at $T=\SI{30}{\micro\kelvin}$. Color scale for the potential is the same as in \autoref{fig:potential}(a).}
		\label{fig:ensemble_evol}
	\end{figure}

	To facilitate comparison and interpretation of the measurement results, we perform Monte-Carlo simulations	of thermal atomic ensembles with 1000 non-interacting classical particles moving in the trap potential, for ensemble temperatures ranging between $\SI{20}{\micro\kelvin}$ and $\SI{200}{\micro\kelvin}$. No external decoherence mechanisms (e.g. fluctuations in magnetic field or trap powers) are taken into account. The motion of the ensemble members is tracked for $\SI{119.5}{\micro\second}$ in steps of $\SI{0.1}{\micro\second}$, and for each step the $(x,y,z)$ position and velocity of each atom are recorded, as is the coupling coefficient to the polarization components of the Raman transfer light. To evaluate the pulse angle for a Raman transfer pulse for individual atoms, the coupling is integrated over the pulse duration. The trap laser powers used for the simulations are $P_{\mathrm{red}}=\SI{1.42}{\milli\watt}$ and $P_{\mathrm{blue}}=\SI{7.7}{\milli\watt}$. Snapshots of the ensemble time evolution after the first pulse of the sequence are illustrated in \autoref{fig:ensemble_evol}.
	
	From the classical simulation data, we can calculate the expected fringe visibility, which lets us compare the experimental data to the simulated ensembles for various temperatures. Simulation results are shown as dashed, dotted and full lines for different ensemble temperatures in \autoref{fig:result_exp_and_sim}.
	
	Comparing to the classical simulation, we find reasonable agreement between the oscillation frequencies and damping times for the data and the simulations for ensemble temperatures between $\SI{25}{\micro\kelvin}$ and $\SI{40}{\micro\kelvin}$. This temperature range fits with previous estimates of the ensemble temperature \cite{Sorensen2016}. However, we also note that the simulations consistently underestimate the depth of the first drop in visibility after half a sloshing period.  
	The overall still reasonable agreement between experimental data and simulation results gives confidence in the simulation as a tool to understand the motion of atomic ensembles at the present ensemble temperature.
	
	When following individual simulated particles with high thermal energy, we observe that the strong anharmonicity of the trap potential leads to particle trajectories only rarely visiting regions close to the nanofiber (see e.g. the two darkest green dots in \autoref{fig:ensemble_evol}). 
	This behavior suggests that good motional averaging of the noisy coupling strength in the ensemble takes far longer than a single trap oscillation \cite{Beguin2018}. 
	
	In a proper quantum mechanical description, despite the cancellation of dipole forces, the localized tagging operation changes the atomic momentum distribution. This can be viewed equivalently as driving motional sideband transitions with the short Fourier broadened Raman pulses. Such dynamics are not taken into account in the current classical model, but will be relevant for colder ensembles with only few initially occupied motional states.
	
	\section{Conclusion}
	
	In summary, we have shown that the thermal motion of an atomic ensemble trapped around a nanofiber can be monitored using short pulses of light which is Stark-shift compensated. The observed oscillation frequencies in the signals match an expected radial trap frequency of $\sim\SI{100}{\kilo\hertz}$. The dephasing of the atomic wave packets leads to a damping of the signal on the time scale of $\SI{10}{\micro\second}$. Corresponding signals can be calculated from simulations of classical particles in the trap potential, at various ensemble temperatures. Comparing data to simulations, we find reasonable agreement for simulated ensemble temperatures between $\SI{25}{\micro\kelvin}$ and $\SI{40}{\micro\kelvin}$. 
	
	For ensemble temperatures that are not low compared to the motional level spacing, a classical point particle model is sufficient to describe atomic motion. With the recent progress towards ground state cooling, it will be relevant and necessary in future work to explicitly take the quantized motion and the back-action of the tagging operation onto the particle momentum into account.
	
	\begin{acknowledgements}
		We gratefully acknowledge funding via the European Research Council grant Quantum-N and by the Villum Foundation. 
	\end{acknowledgements}
	
	\bibliographystyle{spphys}


\begin{thebibliography}{10}
\providecommand{\url}[1]{{#1}}
\providecommand{\urlprefix}{URL }
\expandafter\ifx\csname urlstyle\endcsname\relax
  \providecommand{\doi}[1]{DOI \discretionary{}{}{}#1}\else
  \providecommand{\doi}{DOI \discretionary{}{}{}\begingroup
  \urlstyle{rm}\Url}\fi

\bibitem{Kien2004}
F.L. Kien, V.I. Balykin, K.~Hakuta, Phys. Rev. A
  \textbf{70}, 063403 (2004).
\newblock \href{https://doi.org/10.1103/PhysRevA.70.063403}{\doi{10.1103/PhysRevA.70.063403}}

\bibitem{Vetsch2010}
E.~Vetsch, D.~Reitz, G.~Sagu{\'{e}}, R.~Schmidt, S.T. Dawkins,
  A.~Rauschenbeutel, Phys. Rev. Lett. \textbf{104}, 203603 (2010).
\newblock \href{https://doi.org/10.1103/PhysRevLett.104.203603}{\doi{10.1103/PhysRevLett.104.203603}}

\bibitem{Lacroute2012}
C.~Lacr{\^{o}}ute, K.S. Choi, A.~Goban, D.J. Alton, D.~Ding, N.P. Stern, H.J.
  Kimble, New J. Phys. \textbf{14}, 023056 (2012).
\newblock \href{https://doi.org/10.1088/1367-2630/14/2/023056}{\doi{10.1088/1367-2630/14/2/023056}}

\bibitem{Beguin2014}
J.B. B{\'{e}}guin, E.M. Bookjans, S.L. Christensen, H.L. S{\o}rensen, J.H.
  M{\"{u}}ller, E.S. Polzik, J.~Appel, Phys. Rev. Lett. \textbf{113}, 263603 (2014).
\newblock \href{https://doi.org/10.1103/PhysRevLett.113.263603}{\doi{10.1103/PhysRevLett.113.263603}}

\bibitem{Gouraud2015}
B.~Gouraud, D.~Maxein, A.~Nicolas, O.~Morin, J.~Laurat, Phys. Rev. Lett.
  \textbf{114}, 180503 (2015).
\newblock \href{https://doi.org/10.1103/PhysRevLett.114.180503}{\doi{10.1103/PhysRevLett.114.180503}}

\bibitem{Kato2015}
S.~Kato, T.~Aoki, Phys. Rev. Lett. \textbf{115}, 093603 (2015).
\newblock \href{https://doi.org/10.1103/PhysRevLett.115.093603}{\doi{10.1103/PhysRevLett.115.093603}}

\bibitem{Li2017}
W.~Li, J.~Du, V.G. Truong, S.~{Nic Chormaic}, Appl. Phys. Lett.
  \textbf{110}, 253102 (2017).
\newblock \href{https://doi.org/10.1063/1.4986789}{\doi{10.1063/1.4986789}}

\bibitem{Ruddell2017}
S.K. Ruddell, K.E. Webb, I.~Herrera, A.S. Parkins, M.D. Hoogerland, Optica
  \textbf{4}(5), 576 (2017).
\newblock \href{https://doi.org/10.1364/optica.4.000576}{\doi{10.1364/optica.4.000576}}

\bibitem{Solano2017}
P.~Solano, P.~Barberis-Blostein, F.K. Fatemi, L.A. Orozco, S.L. Rolston, Nat.
  Commun. \textbf{8}, 1857 (2017).
\newblock \href{https://doi.org/10.1038/s41467-017-01994-3}{\doi{10.1038/s41467-017-01994-3}}

\bibitem{LeKien2016}
F.~{Le Kien}, A.~Rauschenbeutel, Phys. Rev. A \textbf{93}, 013849 (2016).
\newblock \href{https://doi.org/10.1103/PhysRevA.93.013849}{\doi{10.1103/PhysRevA.93.013849}}

\bibitem{Beguin2018}
J.B. B{\'{e}}guin, J.H. M{\"{u}}ller, J.~Appel, E.S. Polzik, Phys. Rev. X
  \textbf{8}, 031010 (2018).
\newblock \href{https://doi.org/10.1103/PhysRevX.8.031010}{\doi{10.1103/PhysRevX.8.031010}}

\bibitem{Kato2019}
S.~Kato, N.~N{\'{e}}met, K.~Senga, S.~Mizukami, X.~Huang, S.~Parkins, T.~Aoki,
  Nat. Commun. \textbf{10}, 1160 (2019).
\newblock \href{https://doi.org/10.1038/s41467-019-08975-8}{\doi{10.1038/s41467-019-08975-8}}

\bibitem{Albrecht2016}
B.~Albrecht, Y.~Meng, C.~Clausen, A.~Dareau, P.~Schneeweiss, A.~Rauschenbeutel,
  Phys. Rev. A \textbf{94}, 061401(R) (2016).
\newblock \href{https://doi.org/10.1103/PhysRevA.94.061401}{\doi{10.1103/PhysRevA.94.061401}}

\bibitem{Ostfeldt2017}
C.~{\O}stfeldt, J.B.S. B{\'{e}}guin, F.T. Pedersen, E.S. Polzik, J.H.
  M{\"{u}}ller, J.~Appel, Opt. Lett. \textbf{42}(21), 4315 (2017).
\newblock \href{https://doi.org/10.1364/ol.42.004315}{\doi{10.1364/ol.42.004315}}

\bibitem{Meng2018}
Y.~Meng, A.~Dareau, P.~Schneeweiss, A.~Rauschenbeutel, Phys. Rev. X
  \textbf{8}, 031054 (2018).
\newblock \href{https://doi.org/10.1103/PhysRevX.8.031054}{\doi{10.1103/PhysRevX.8.031054}}

\bibitem{Oblak2008}
D.~Oblak, J.~Appel, P.J. Windpassinger, U.B. Hoff, N.~Kj{\ae}rgaard, E.S.
  Polzik, Eur. Phys. J. D \textbf{50}(1), 67 (2008).
\newblock \href{https://doi.org/10.1140/epjd/e2008-00192-1}{\doi{10.1140/epjd/e2008-00192-1}}

\bibitem{Sorensen2016}
H.L. S{\o}rensen, J.B. B{\'{e}}guin, K.W. Kluge, I.~Iakoupov, A.S. S{\o}rensen,
  J.H. M{\"{u}}ller, E.S. Polzik, J.~Appel, Phys. Rev. Lett. \textbf{117}, 133604 (2016).
\newblock \href{https://doi.org/10.1103/PhysRevLett.117.133604}{\doi{10.1103/PhysRevLett.117.133604}}

\end{thebibliography}

\end{document}